\documentclass[iop]{emulateapj}

\slugcomment{Submitted to JAE}

\shorttitle{Precise Pointing \& Stabilization of BIT}
\shortauthors{Romualdez et al.}

\usepackage{amsmath}
\usepackage{color}
\usepackage{soul}
\usepackage{graphicx}
\usepackage{fancyhdr}
\usepackage{accents}
\usepackage{booktabs}
\usepackage{amsthm,amssymb,url}
\usepackage{longtable}
\usepackage[T1]{fontenc}
\usepackage{savesym}
\savesymbol{tablenum}
\usepackage{siunitx}
\restoresymbol{SIX}{tablenum}
\newcommand{\coord}[1]{\underaccent{\vec}{\mathcal{F}}_{#1}}
\newcommand{\changed}[1]{\textcolor{black}{#1}}

\begin{document}

%\runninghead{Romualdez et al.}

\title{Precise Pointing and Stabilization Performance for the Balloon-borne Imaging Testbed (BIT): 2015 Test Flight}

\author{L. J. Romualdez\altaffilmark{1}, P. Clark\altaffilmark{3}, C. J. Damaren\altaffilmark{1}, M. N. Galloway\altaffilmark{2}, \\J. W. Hartley\altaffilmark{2}, L. Li\altaffilmark{1}, R. J. Massey\altaffilmark{2}, C. B. Netterfield\altaffilmark{2}}

%% Notice that each of these authors has alternate affiliations, which
%% are identified by the \altaffilmark after each name.  Specify alternate
%% affiliation information with \altaffiltext, with one command per each
%% affiliation.

\altaffiltext{1}{Institute for Aerospace Studies, University of Toronto, Canada}
\altaffiltext{2}{Department of Physics, University of Toronto, Canada}
\altaffiltext{3}{Centre for Advanced Instrumentation, University of Durham, United Kingdom}

%\author{L. Javier Romualdez\affilnum{1}, Paul Clark\affilnum{3}, Christopher J. Damaren\affilnum{1}, Mathew N. Galloway\affilnum{2}, John W. Hartley\affilnum{2}, Lun Li\affilnum{1}, Richard J. Massey\affilnum{2}, C. Barth Netterfield\affilnum{2}}

%\corrauth{L. Javier Romualdez, Institute for Aerospace Studies, University of Toronto, 4925 Dufferin Street, Toronto, Ontario, M3H 5T6, Canada}
%\email{javier.romualdez@mail.utoronto.ca}

\begin{abstract}
Balloon-borne astronomy offers an attractive option for experiments that require precise pointing and attitude stabilization, due to a large reduction in the atmospheric interference observed by ground-based systems as well as the low-cost and short development time-scale compared to space-borne systems. The Balloon-borne Imaging Testbed (BIT) is an instrument designed to meet the technological requirements of high precision astronomical missions and is a precursor to the development of a facility class instrument with capabilities similar to the Hubble Space Telescope. The attitude determination and control systems (ADCS) for BIT, the design, implementation, and analysis of which are the focus of this paper, compensate for compound pendulation effects and other sub-orbital disturbances in the stratosphere to within 1-2$^{\prime\prime}$ (rms), while back-end optics provide further image stabilization down to 0.05$^{\prime\prime}$ (not discussed here). During the inaugural test flight from Timmins, Canada in September 2015, BIT ADCS pointing and stabilization performed exceptionally, with coarse pointing and target acquisition to within < 0.1$^\circ$ and fine stabilization to 0.68$^{\prime\prime}$ (rms) over long (10-30 minute) integrations. This level of performance was maintained during flight for several tracking runs that demonstrated pointing stability on the sky for more than an hour at a time. To refurbish and improve the system for the three-month flight from New Zealand in 2018, certain modifications to the ADCS need to be made to smooth pointing mode transitions and to correct for internal biases observed during the test flight. Furthermore, the level of autonomy must be increased for future missions to improve system reliability and robustness.
\end{abstract}


\maketitle

\section{Background}

\subsection{Overview of balloon-borne astronomy}
\label{ss:overview_bb_astro}

For many astronomical and astrophysical experiments, scientific balloon-borne payloads offer an attractive trade-off between space-borne systems, which are often expensive and require a long development time-scale, and ground-based systems, which suffer greatly from atmospheric effects or ``seeing'' on the order of 1$^{\prime\prime}$ (arcsecond). Specifically, certain dark matter and dark energy related missions, as proposed by the Canadian 2010 Long Range Plan for Astronomy (LRP2010) and the US Astro2010 panel \citep{RefWorks:1}, require systems with a highly precise pointing resolution over large integration times in order to detect photometric redshifts in supernovae for lensing studies involved with such experiments \citep{RefWorks:1}. At an altitude of approximately 35-40 km, balloon-borne instrumentation provides a viable platform for meeting these requirements since a theoretical resolution of 0.01$^{\prime\prime}$ can be achieved at near-UV and visible wavelengths (300-900 \si\micro\si\metre). 

\begin{figure*}[ht]
\centering
\includegraphics[width=0.8\textwidth]{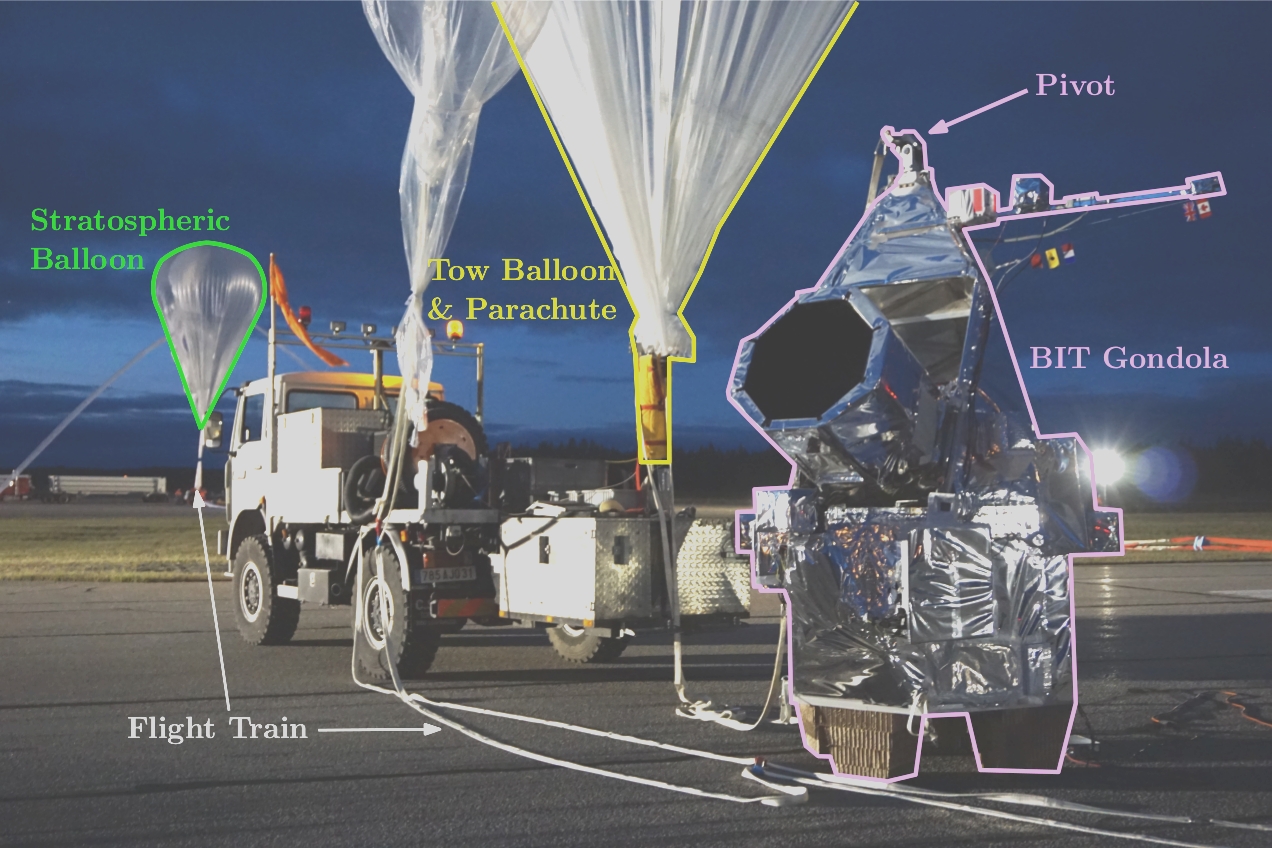}
\caption[numbered]{The Balloon-borne Imaging Testbed (BIT) approximately one hour before the September 2015 test launch from Timmins, Canada; the stratospheric balloon (left in the distance) is attached to the flight train (along the bottom) via a smaller tow balloon and parachute (attached to truck) with the BIT payload/gondola (on the right).}
\label{fig:BIT_launch}
\end{figure*}

From a control design perspective, the operational environment for a high precision pointing balloon-borne instrument must be given special consideration, since any payload operating in the stratosphere is subject to a number of sub-orbital effects that are not present in a space or ground-based environment. A stratospheric launch vehicle is typically divided into three main structures (see Figure \ref{fig:BIT_launch}): a 1 million cubic metre helium balloon, a 60-100 m long \textit{flight train} containing the return parachute, and the payload or \textit{gondola} housing all scientific equipment. From a disturbance point-of-view at float altitude, gravity-driven effects dominate the low frequency regime and are manifested as a six degree-of-freedom double pendulative motion about the balloon-to-flight-train connection ($\sim $ 0.01-0.05 Hz) as well as about the flight-train-to-payload or \textit{pivot} connection ($\sim $ 0.5-1 Hz) \citep{RefWorks:28}. Despite the lack of atmosphere ($\sim$ 3 mbar), the stratosphere both directly and indirectly affects the balloon-borne payload via stratospheric wind-shears causing intermittent translational and rotational acceleration ($\sim $ 0.5 g) as well as a slowly varying balloon rotation ($\sim $ 0.03 rpm) \citep{RefWorks:1,RefWorks:28}. For remote sensing missions that require long integration times ($\sim $ 5-10 min) such as those proposed above, effects due to the rotation of the Earth and the precession of the balloon during integration must also be taken into account. 
 
Historically, the most notable experiment in the field of high pointing precision astronomical systems from the stratosphere is Stratoscope II, a balloon-borne visible range telescope which pioneered advances and innovations in atmospheric and space astronomy from 1967-1973 \citep{RefWorks:95,RefWorks:88}. Being one of the first sub-orbital telescopes, post-flight analysis of Stratoscope II data showed that it was capable of a 0.02$^{\prime\prime}$ focal plane equivalent pointing stability with an image resolution of 0.2$^{\prime\prime}$ for integration periods upwards of 1 minute, which was unmatched by any other instrument in use at the time \citep{RefWorks:95, RefWorks:88}. To accomplish this, Stratoscope II had a two-stage pointing and stabilization scheme: a coarse routine that stabilized the instrument to 15$^{\prime\prime}$ using a combination of stepper motors and smooth torquers (i.e. direct-drive motors), and a fine routine that used a 4-12 Hz bandwidth transfer lens to track out remaining disturbances on the image plane \citep{RefWorks:88}. With this control methodology, Stratoscope II set the precedent for future high precision astronomical instrumentation, and, as such, represents the state-of-the-art for control design and pointing stability of balloon-borne imaging telescopes to date. 

It should be noted that Stratoscope II relied primarily on a low-gain control scheme, where disturbances from the balloon-borne environment are passively controlled the majority of the time via balancing and occasionally corrected to compensate for larger disturbances and coupling \citep{RefWorks:88}. Consequently, significant image processing in post-flight analysis was required to demonstrate the quoted pointing stability and image resolution for only a handful of images \citep{RefWorks:88}. As a result, there is precedent to develop stratospheric instruments that can achieve a high degree of pointing stabilization and image resolution on demand with a high-gain closed loop system that actively and continuously corrects for disturbances. The Balloon-borne Imaging Testbed (BIT) is such an instrument that attempts to meet these technological requirements, where the design, flight implementation, and performance of on-board, real-time attitude determination and control systems (ADCS) is the focus of this work.

\subsection{The Balloon-borne Imaging Testbed (BIT)}
\label{ss:bb_enviro}

The Balloon-borne Imaging Testbed (BIT) is a joint project between the University of Toronto Astrophysics Department (U of T) as well as the Institute for Aerospace Studies (UTIAS), the Durham University Centre for Advanced Instrumentation (CfAI), and the Jet Propulsion Laboratory (JPL-NASA). The overall objective of the project is to develop, build, and test a balloon-borne telescope platform designed to meet the technological requirements of astronomical missions that require a high degree of pointing accuracy and stabilization in a way that is ``better, faster, and cheaper'' than similar space-borne or ground-based telescope missions. As such, the BIT project is structured to develop design-oriented and implementation methodologies for generic high-precision balloon-borne astronomical missions while demonstrating an overall system pointing resolution of 0.05$^{\prime\prime}$, a capability that is second only to the Hubble Space Telescope (HST). The pointing stabilization and control results of the September 2015 flight from Timmins, Canada with the Canadian Space Agency (CSA) and Centre national d'etudes spatiales (CNES) are presented in this work. The pointing specifications will be further verified during a 24 hour performance flight from Fort Sumner, New Mexico as well as a three month fully operational flight in 2018 from New Zealand, which will demonstrate the capabilities of BIT as a facility class instrument on an ultra long-duration balloon flight (ULDB). 

\begin{figure*}[ht]
\centering
\includegraphics[width=0.8\textwidth,page=1]{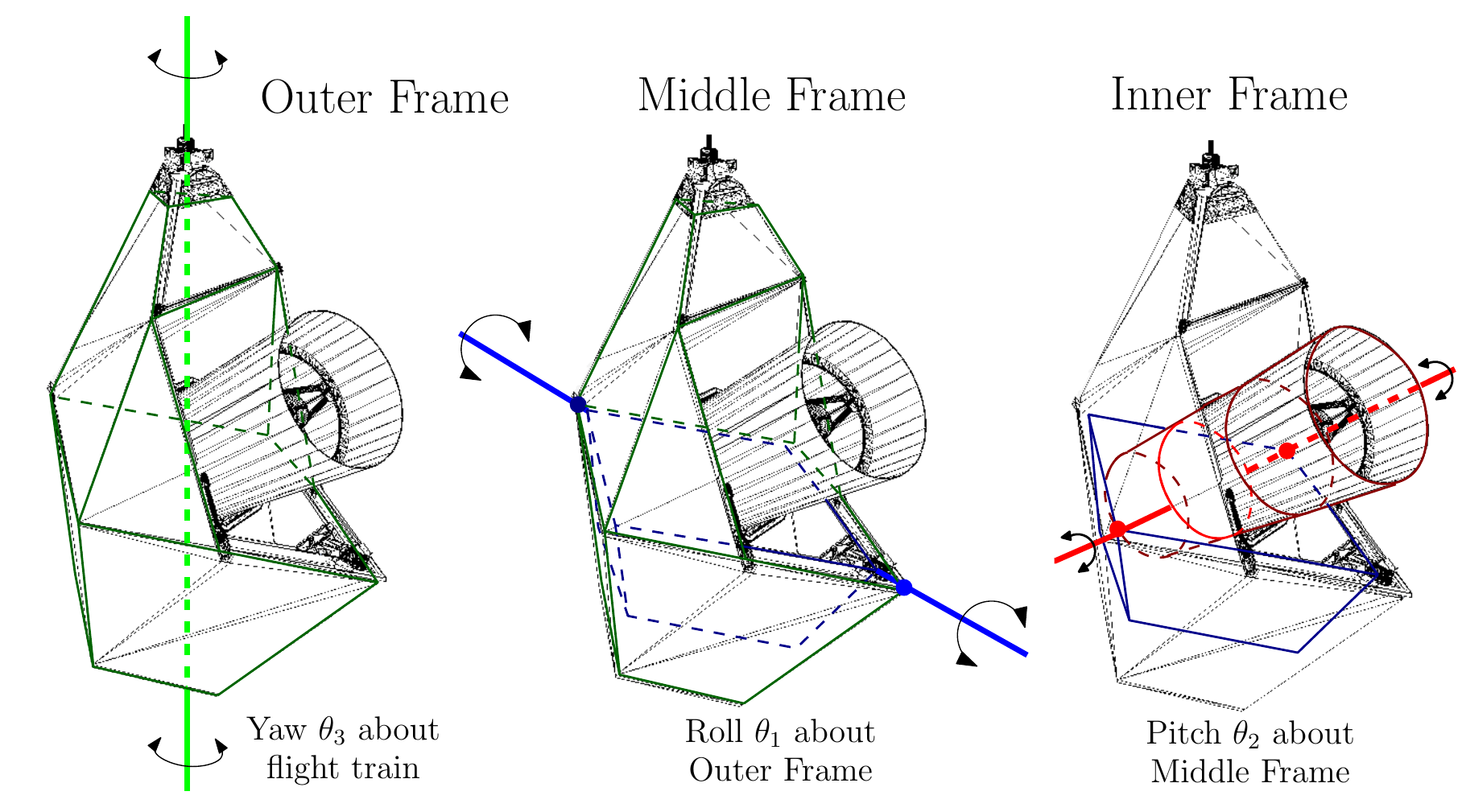}
\caption[numbered]{Schematic diagram of the BIT gondola, comprised of an outer frame (left), middle frame (centre), and inner frame (right); gimbal coordinates are defined by $\boldsymbol\theta = [\theta_1 \ \theta_2 \ \theta_3 ]^T$ as a 3-1-2 Euler sequence about their respective axes (thick line); yaw $\theta_3$ is unconstrained whereas roll $\theta_1$ and pitch $\theta_2$ have gimbal ranges of $\pm 6^\circ$ and  20-57$^\circ$, respectively.}
\label{fig:BIT_phys_arch}
\end{figure*}

As shown in Figure \ref{fig:BIT_phys_arch}, the physical architecture of the BIT gondola can be broken down into three main gimballed substructures or frames, each actuated about their own orthogonal axis. The outer frame structure is actuated about the yaw $\theta_3$ axis via a reaction wheel located at the base and co-actuated at the pivot connection to prevent flight train twisting and reaction wheel saturation. The middle frame is actuated relative to the outer frame about the roll $\theta_1$ axis and is physically constrained by a $\pm 6^\circ$ range. Similarly, the inner frame, which contains the telescope, optics, and the corresponding flight electronics, is actuated relative to the middle frame about the pitch $\theta_2$ axis and has a full range from 20-57$^\circ$ relative to horizontal. Using these three gimballed frames, the attitude of the telescope on the inner frame relative to targets on the sky is fully controlled with hardware specified to stabilize the telescope to within 1-2$^{\prime\prime}$. Back-end optics located at the rear of the telescope further stabilize the image on the telescope focal plane to 0.05$^{\prime\prime}$ according to the desired specifications. The control of the inner frame attitude and telescope stabilization down to 1-2$^{\prime\prime}$ is the primary focus of this work, whereas the performance of the back-end optics is beyond the scope of this paper.

\section{BIT ADCS Design}

\subsection{Attitude determination}
\label{sss:att_determ}

\subsubsection{Celestial coordinate systems.}
\label{sss:cel_coord:syst}

Given the astronomy- and cosmology-related applications of BIT, all attitude determination and state estimation is done in the Earth-centric equatorial frame $\coord{E}$, which is based in right ascension ($RA$), declination ($Dec$), and field rotation ($FR$) coordinates \citep{RefWorks:89}. As such, the orientation of the telescope body frame $\coord{b}$ with respect to the equatorial inertial frame $\coord{E}$ is represented by the 3-2-1 Euler sequence
\begin{eqnarray}
\mathbf{C}_{bE} = \coord{b}\cdot\coord{E}^T = \mathbf{C}_x(FR)\mathbf{C}_y(-Dec)\mathbf{C}_z(RA)
\label{eq:equat_coords}
\end{eqnarray}
where $\mathbf{C}_x(\cdot)$, $\mathbf{C}_y(\cdot)$, and $\mathbf{C}_z(\cdot)$ are the respective elementary rotations about the $x$, $y$, and $z$ axes. Note that the signs for the coordinates are based on astronomical conventions.

Although all attitude determination takes place in the equatorial frame, it is often useful to reference the local horizontal frame $\coord{H}$, which is fixed to and rotates with the Earth, since certain sensors provide attitude information that is intrinsically referenced to an Earth-fixed frame. The coordinates of $\coord{b}$ with respect to the horizontal frame $\coord{H}$ are based in azimuth ($Az$), elevation ($El$), and image rotation ($IR$) \citep{RefWorks:89}, which parameterize the 3-2-1 Euler sequence
\begin{eqnarray}
\mathbf{C}_{bH} = \coord{b}\cdot\coord{H}^T  = \mathbf{C}_x(-IR)\mathbf{C}_y(-El)\mathbf{C}_z(Az)
\label{eq:horiz_coords}
\end{eqnarray}
where again the signs are based on astronomical conventions.

Since $\coord{H}$ is Earth-fixed and rotates with respect to $\coord{E}$, the orientation of $\coord{H}$ with respect to $\coord{E}$ can be given in terms of local latitude $\phi_\ell$, local longitude $\psi_\ell$, and local sidereal time $\theta_\ell$ \citep{RefWorks:89} as
\begin{eqnarray}
\mathbf{C}_{HE} = \coord{H}\cdot\coord{E}^T = \mathbf{C}_y(\phi_\ell-\tfrac{\pi}{2})\mathbf{C}_z(-(\psi_\ell+\theta_\ell+\pi))
\label{eq:equat_to_horiz}
\end{eqnarray}
where it is clear that $\mathbf{C}_{HE} = \mathbf{C}_{HE}(t)$ since sidereal time is based on UTC time. From this, all pertinent information regarding telescope orientation in either $\coord{E}$ or $\coord{H}$ is captured in a way that is easily applied to attitude determination schemes involving rotation matrix estimation, such as those presented here.

\subsubsection{State model and prediction.}
\label{sss:state_mod_pred}

In general, the rotational kinematics of a rigid body can be shown to have the following discrete-time form \citep{RefWorks:90}:
\begin{eqnarray}
\mathbf{C}_{bE,k} = \boldsymbol\Psi_{k}\mathbf{C}_{bE,k-1}
\label{eq:discrete_rot_kin}
\end{eqnarray}
where 
\begin{eqnarray}
\boldsymbol\Psi_{k} &=& \cos(\psi_k)\mathbf{1}+(1-\cos(\psi_k))\left(\frac{\boldsymbol\psi_k}{\psi_k}\right)\left(\frac{\boldsymbol\psi_k}{\psi_k}\right)^T+\nonumber\\
&&...-\sin(\psi_k)\left(\frac{\boldsymbol\psi_k}{\psi_k}\right)^\times
\label{eq:rot_kin_psi}\\
\boldsymbol\psi_k &=& \boldsymbol\omega_{b,k}T_k, \ \psi_k = \|\boldsymbol\psi_k\|\nonumber
\end{eqnarray}
Here, $\mathbf{C}_{bE,k}$ is the orientation of the telescope $\coord{b}$ with respect to equatorial frame $\coord{E}$, and $T_k$ is the sampling period for the discrete time system, where it is noted that the construction of $\boldsymbol\Psi_k$ preserves the orthogonality of $\mathbf{C}_{bE}$ at each time index $k$. The angular velocity $\boldsymbol\omega_{b,k}$ expressed in $\coord{b}$ is measured by three orthogonal single-axis KVH\textsuperscript{\textregistered} DSP-1750 fibre-optic rate gyroscopes (sensor characteristics are given in Table \ref{table:sensor_char}) according to the sensor model
\begin{eqnarray}
\boldsymbol\omega_b = \mathbf{A}_{bg}(\boldsymbol\omega_g+\mathbf{b}_g)
\label{eq:total_gyro_calib}
\end{eqnarray}
where $\boldsymbol\omega_g$ is the raw measurement, $\mathbf{b}_g$ is the rate gyroscope measurement bias, and $\mathbf{A}_{bg}$ is a calibration matrix that accounts for orthogonality and scale factor misalignments.

Using the fact that a small perturbation of a rotation matrix is, in general, given by $(\mathbf{1}-\delta\boldsymbol\phi^\times)$ \citep{RefWorks:90}, the error kinematics for the discrete system can be shown to have the following linear form:
\begin{eqnarray}
\underbrace{
\begin{bmatrix}
\delta\boldsymbol\phi_{bE,k}\\
\delta\mathbf{b}_{g,k}
\end{bmatrix}
}_{\delta\mathbf{x}_k} &=&
\underbrace{
\begin{bmatrix}
\boldsymbol\Psi_k & \mathbf{A}_{bg}T_k\\
\mathbf{0} & \mathbf{1}
\end{bmatrix}
}_{\mathbf{H}_{x,k}}
\underbrace{
\begin{bmatrix}
\delta\boldsymbol\phi_{bE,k-1}\\
\delta\mathbf{b}_{g,k-1}
\end{bmatrix}
}_{\delta\mathbf{x}_{k-1}}+\nonumber\\
&&...+\underbrace{
\begin{bmatrix}
\mathbf{A}_{bg}T_k & \mathbf{0}\\
\mathbf{0} & \mathbf{1}
\end{bmatrix}
}_{\mathbf{H}_{w,k}}
\underbrace{
\begin{bmatrix}
\delta\boldsymbol\omega_g\\
\delta\boldsymbol\beta_g
\end{bmatrix}
}_{\mathbf{w}_k}
\label{eq:gyro_bias_state_eq}
\end{eqnarray}
Here, $\delta\boldsymbol\omega_g$ represents the noise model for the raw rate gyroscope measurements and $\delta\boldsymbol\beta_g$ models the drift in the rate gyroscope bias as a random walk process. It is clear that the matrices $\mathbf{H}_{x,k}$ and $\mathbf{H}_{w,k}$ are the state and measurement Jacobians, respectively, for the predictive step of an extended Kalman filter (EKF) \citep{RefWorks:40}. 

\subsubsection{Measurement model and correction.}
\label{sss:meas_mod_and_corr}

External measurements of the telescope attitude are obtained from two separate sources: coarse sensors, which include three-axis magnetometer and optical encoder measurements, and star cameras. For large slews and coarse pointing up to $\sim 0.5^\circ$, the coarse sensors are the primary external measurement for attitude determination, whereas for pointing stabilization and fine pointing up to $< 1$-$2^{\prime\prime}$, the star cameras are the dominant external measurement.

For coarse sensors, the telescope orientation can be estimated to within the $\sim 1$-$2^{\prime}$ (arcminute) pendulations of the outer frame based on the gimbal position of the three frames. As a result, the orientation of the telescope with respect to $\coord{H}$ can be estimated by from a 3-1-2 Euler sequence $\mathbf{C}_{IF,OF}$ from the outer frame to the inner (telescope) using coarse sensors as well as a calibration term $\mathbf{C}_{cal,coarse}$ to account for the offset of the outer frame with respect to the horizon due to mass imbalances. Thus, the telescope orientation with respect to  $\coord{E}$ is given by
\begin{eqnarray}
\mathbf{C}_{bE,meas,coarse} &=& \mathbf{C}_{IF,OF}\mathbf{C}_{cal,coarse}\mathbf{C}_{HE}(t)
\end{eqnarray}
where
\begin{eqnarray}
\mathbf{C}_{IF,OF} &=& \mathbf{C}_y(\theta_2)\mathbf{C}_x(\theta_1)\mathbf{C}_z(\theta_3)\nonumber
\label{eq:312_gimbal}
\end{eqnarray}
and $\mathbf{C}_{HE}(t)$ is given by (\ref{eq:equat_to_horiz}).

The roll gimbal angle $\theta_1$ of the middle frame with respect to the outer frame is obtained from a 16-bit absolute optical encoder from BEI Sensors (further sensor details are given in Table \ref{table:sensor_char}). A combination of encoder measurements on the inner frame axis and pitch stepper motor counts provide the pitch angle $\theta_2$ of inner (telescope) frame with respect to the middle frame. Lastly, the yaw gimbal angle $\theta_3$ of the outer frame with respect to due north is obtained from a Honeywell HMR2300 three-axis magnetometer (details in Table \ref{table:sensor_char}) according to the simple calibrated model
\begin{eqnarray}
\theta_3 = \mbox{atan2}(s_x(m_x-d_x),s_y(m_y-d_y))+\theta_{3,0}
\label{eq:maggie_model}
\end{eqnarray}
where $m_{x,y}$ are the raw magnetometer measurements along the $x$ (roll) and $y$ (pitch) axes. The terms $s_{x,y}$ and $d_{x,y}$ are scale factor and offset calibration terms determined experimentally with $\theta_{3,0}$ as the overall yaw offset term.

From this estimate of the telescope orientation $\mathbf{C}_{bE}$, the measurement model takes on the trivial form 
\begin{eqnarray}
\mathbf{C}_{bE,meas,coarse,k} = (\mathbf{1}-\delta\mathbf{n}_{coarse})\mathbf{C}_{bE,k}
\label{eq:coarse_meas_model}
\end{eqnarray}
where $\delta\mathbf{n}_{coarse}$ is a measurement noise term that models the accuracy of the coarse sensors. From this, it can be shown that the measurement and noise Jacobians for the correction step of an EKF \citep{RefWorks:40} are given by $\mathbf{G}_{x,k} = \mathbf{1}$ and $\mathbf{G}_{n,k} = \mathbf{1}$, respectively. Similarly, the innovation or error term $\mathbf{e}_k$ is given by
\begin{eqnarray}
\mathbf{e}_k^\times \approx \mathbf{1}-\mathbf{C}_{bE,meas,k}\mathbf{C}_{bE,k}^T
\label{eq:ekf_innov}
\end{eqnarray}
which, with a suitable gain $\mathbf{K}_k$, can be used to correct the telescope orientation $\mathbf{C}_{bE,k}^-$ and rate gyroscope bias $\mathbf{b}_{g,k}^-$ from the prediction step according to
\begin{eqnarray}
\mathbf{C}_{bE,k} = \boldsymbol\Xi_k\mathbf{C}_{bE,k}^-\\
\mathbf{b}_{g,k} = \mathbf{b}_{g,k}^-+\delta\boldsymbol\xi_{g,k}\nonumber
\label{eq:coarse_rot_correction}
\end{eqnarray}
where
\begin{eqnarray}
\boldsymbol\Xi_{k} &=& \cos(\delta\xi_{bE,k})\mathbf{1}+\nonumber\\
&&...+(1-\cos(\delta\xi_{bE,k}))\left(\frac{\delta\boldsymbol\xi_{bE,k}}{\delta\xi_{bE,k}}\right)\left(\frac{\delta\boldsymbol\xi_{bE,k}}{\delta\xi_{bE,k}}\right)^T+\nonumber\\
&&...-\sin(\delta\xi_{bE,k})\left(\frac{\delta\boldsymbol\xi_{bE,k}}{\delta\xi_{bE,k}}\right)^\times
\label{eq:coarse_rot_correction_phi}
\end{eqnarray}
and
\begin{eqnarray}
\delta\xi_{bE,k} &=& \|\delta\boldsymbol\xi_{bE,k}\|\nonumber\\
\begin{bmatrix}
\delta\boldsymbol\xi_{bE,k}\\ \delta\boldsymbol\xi_{g,k}
\end{bmatrix} &=&
\mathbf{K}_k\mathbf{e}_k\nonumber
\end{eqnarray}
Note that similar to (\ref{eq:rot_kin_psi}), this formulation preserves rotation matrix orthogonality.

For pointing stabilization and fine pointing, two star cameras are used to track the sky and mitigate disturbances: the bore star camera along the boresight of the telescope, which provides fine star camera measurement $\mathbf{C}_{bE,meas,sc} = \mathbf{C}_{bE,meas,bore}$ since it is aligned with $\coord{b}$ ($\mathbf{C}_{bE,meas,bore}$ from bore star camera coordinates), and the roll star camera, which provides $\mathbf{C}_{bE,meas,sc} = \mathbf{C}_{rb}\mathbf{C}_{bE,meas,roll}$ since it is orthogonal to the telescope boresight ($\mathbf{C}_{rb}$ from calibration and $\mathbf{C}_{bE,meas,roll}$ from roll star camera coordinates). The measurement model and correction for the two star cameras are analogous to the coarse measurement model given in (\ref{eq:coarse_meas_model}) except for, of course, the magnitude of the noise terms contributing to the pointing solution uncertainty (i.e. $\delta\mathbf{n}_{sc} \ll \delta\mathbf{n}_{coarse}$). In addition, to ensure dominance of star camera contribution to the pointing solution, the variance on the noise term $\delta\mathbf{n}_{coarse}$ is increased during pointing stabilization. Furthermore, to improve coarse attitude determination accuracy, the measurement model given by (\ref{eq:312_gimbal}) is trimmed to correct for outer frame imbalances by computing $\mathbf{C}_{cal,coarse}$ from
\begin{eqnarray}
\mathbf{C}_{cal,coarse} = \mathbf{C}_{IF,OF}^T\mathbf{C}_{bE,meas,sc}\mathbf{C}_{HE}^T(t)
\label{eq:coarse_trim}
\end{eqnarray}

When star cameras provide a full pointing solution (a.k.a. \textit{lost-in-space} mode), the measured equatorial coordinates are converted to a pointing measurement using (\ref{eq:equat_coords}) to give $\mathbf{C}_{bE,meas,sc}$. However, once the initial lock is obtained, the brightest star in each star camera is reported and attitude determination switches to \textit{differential mode} in which only the centroid locations of the brightest blobs are used for feedback. For both star cameras, the centroid pixel coordinates $(x_{sc},y_{sc})$ are estimated as a constant vector in $\coord{E}$ as
\begin{eqnarray}
\mathbf{v}_E = \mathbf{C}_{bE,meas,sc}^T\mathbf{v}_b \approx \mathbf{C}_{bE,meas,sc}^T\begin{bmatrix}
\tfrac{1}{p} \\ x_{sc} \\ y_{sc}
\end{bmatrix}
\label{eq:centroid_model}
\end{eqnarray}
where $p$ is the pixel scale (in rad/px) of the given star camera. From this, the measurement model when using centroids in differential mode is given by
\begin{eqnarray}
\mathbf{v}_{b,meas} = \mathbf{C}_{bE,k}\mathbf{v}_E+\delta\mathbf{n}_{centroid}
\label{eq:centroid_meas_model}
\end{eqnarray}
which can be shown to have the following linear error dynamics:
\begin{eqnarray}
\delta\mathbf{v}_{b,meas}
&=&
\underbrace{
\begin{bmatrix}
(\mathbf{C}_{bE,k}\mathbf{v}_E)^\times & \mathbf{0}
\end{bmatrix}
}_{\mathbf{G}_{x,k}}
\underbrace{
\begin{bmatrix}
\delta\boldsymbol\phi_{bE,k}\\
\delta\mathbf{b}_{g,k}
\end{bmatrix}
}_{\delta\mathbf{x}_{k}}+\nonumber\\
&&...+\delta\mathbf{n}_{centroid}
\label{eq:centroid_linear_model}
\end{eqnarray}
As before, $\mathbf{G}_{x,k}$ and $\mathbf{G}_{n,k}=\mathbf{1}$ are the measurement and noise Jacobians for the correction step of a typical EKF \citep{RefWorks:40}. It is worth noting that each star camera contributes centroid information to the pointing solution asynchronously, so although a single centroid measurement from a single star camera does not provide full attitude information (i.e. no information about the roll axis of the centroid), the contribution of centroid information from both star cameras over time constrains the full pointing estimate.

\subsection{Attitude stabilization and control}
\label{ss:att_stab_cont}

\subsubsection{Target command coordinates.}
\label{sss:targ_comm_coord}

When commanding a desired astronomical target on the sky, coordinates are conventionally specified in the equatorial frame $\coord{E}$ ($RA$ and $Dec$) or in the horizontal frame $\coord{H}$ ($Az$ and $El$) \citep{RefWorks:89}. In both cases, only two of the three coordinates in either frame are specified, which allows for an extra degree of freedom in the pointing specification. Due to physical gimbal constraints, a natural pointing specification would be to select initial gimbal angles that maximize exposure on the sky. Thus, the roll of the middle frame with respect to the outer frame $\theta_1$ was selected for the third pointing coordinate since it is the most restrictive ($\pm 6^\circ$). As such, from (\ref{eq:horiz_coords}) and (\ref{eq:312_gimbal}), the gimbal angles $\boldsymbol\theta = [\theta_1 \ \theta_2 \ \theta_3 ]^T$ are related to the equatorial coordinates simply by
\begin{eqnarray}
\mathbf{C}_x(-IR)\mathbf{C}_y(-El)\mathbf{C}_z(Az) = \mathbf{C}_{IF,OF}\mathbf{C}_{cal,coarse}
\label{eq:horiz_to_gimbal}
\end{eqnarray}
where the left side is a 3-2-1 Euler rotation and the right side is a 3-1-2 Euler rotation. Therefore, given a set ($Az$, $El$, $\theta_1$) as well as a calibrated $\mathbf{C}_{cal,coarse}$, one can find a unique pitch angle $\theta_2$, yaw angle $\theta_3$, and image rotation $IR$ that maximizes the roll gimbal angle range while tracking the sky. Similarly, given coordinates ($RA$, $Dec$, $\theta_1$) with (\ref{eq:equat_coords}) and (\ref{eq:equat_to_horiz}), a corresponding gimbal set $\theta_2$ and $\theta_3$ as well as field rotation $FR$ can be found. 

Additionally, it is also useful for direct gimbal control to project the rate gyroscope measurements to the corresponding gimbal axes. This is done using the mapping matrix $\mathbf{S}_\theta$, which is defined for a 3-1-2 Euler rotation as \citep{RefWorks:90}
\begin{eqnarray}
\boldsymbol\omega_b = \underbrace{\begin{bmatrix}
\cos(\theta_2) & 0 & -\cos(\theta_1)\sin(\theta_2)\\
0 & 1 & \sin(\theta_1)\\
\sin(\theta_2) & 0 & \cos(\theta_1)\cos(\theta_2)
\end{bmatrix}}_{\mathbf{S}_\theta}\underbrace{\begin{bmatrix}
\dot{\theta}_1 \\ \dot{\theta}_2 \\ \dot{\theta}_3
\end{bmatrix}}_{\dot{\boldsymbol\theta}}
\label{eq:312_mapping_matrix}
\end{eqnarray}
In this way, given rate gyroscope measurements $\boldsymbol\omega_b$ and measurements of roll $\theta_1$ and pitch $\theta_2$, the corresponding gimbal rate $\dot{\boldsymbol\theta}$ can be found. This of course is only valid away from the matrix singularity $\theta_1 = \pm\tfrac{\pi}{2}$ \citep{RefWorks:90}, which is a reasonable constraint since the physical bounds of the roll axis are limited to $\pm 6^\circ$.

\subsubsection{Coarse gimbal control.}
\label{sss:coarse_gimb_cont}

Coarse gimbal control is primarily used on BIT for coarse stabilization after launch as well as for coarse target acquisition that requires large slews. As such, this mode of coarse control does not track pendulations of the gondola or the rotation of the sky, but instead sets the local yaw $\theta_3$, pitch $\theta_2$, and roll $\theta_1$ angles to correspond with a given celestial target, as in (\ref{eq:horiz_to_gimbal}). With this method, coarse gimbal control acquires its target nominally to within $<0.1^\circ$.

\begin{figure*}[h]
\centering
\includegraphics[width=0.85\textwidth]{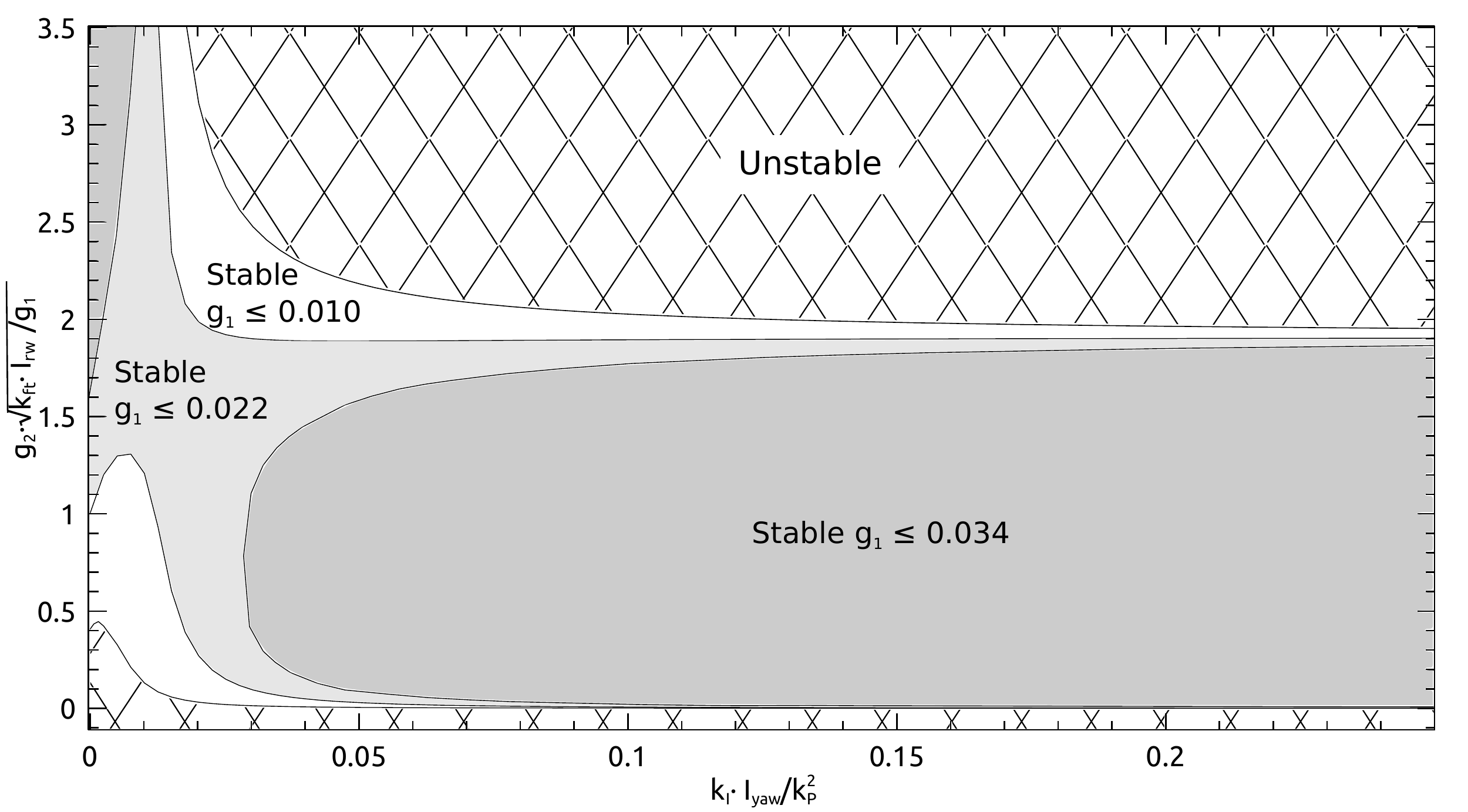}
\caption[numbered]{Stability regions (white and shaded) for momentum dumping and coarse yaw speed control using fully coupled yaw dynamics (unstable region is hatched); gains $g_2$ on the $y$-axis and $k_I$ on the $x$-axis are normalized for a given $g_1$ and $k_P$, respectively, where $k_I\cdot I_{yaw}/k_P^2 \leq 0.25$ is the condition for critical to near-overdamped control from (\ref{eq:rw_control}) assuming uncoupled yaw dynamics; three stability regions are given for $g_1\leq 0.01$ (white and shaded), $g_1\leq 0.022$ (light and dark shaded), and $g_1\leq 0.034$ (dark shaded), where it is clear that increasing the bound on $g_1$ shrinks the stability region.}
\label{fig:yaw_stab_diag}
\end{figure*}

For coarse stabilization and pointing in yaw, the outer frame of the gondola is actuated primarily by a large 20 kg$\cdot$m\textsuperscript{2} reaction wheel driven by a 20 N$\cdot$m Parker frameless DC motor fixed to the bottom of the gondola (see Table \ref{table:actuator_char} for details). To prevent reaction wheel saturation, a speed controlled Applied Motion Products HT17-075 stepper motor actuates the pivot at the top of the outer frame (see Figure \ref{fig:BIT_launch}) through a 100:1 gear reducer. In this way, momentum can be dumped from the reaction wheel through the flight train to the stratospheric balloon. With a prescribed bias reaction wheel speed $\omega_{rw,d}$, the momentum dumping law chosen \citep{RefWorks:85} for yaw control and to prevent reaction wheel saturation is
\begin{eqnarray}
\omega_{piv} = g_1(\omega_{rw}-\omega_{rw,d})+g_2\tau_{rw,comm}
\label{eq:momentum_dump}
\end{eqnarray}
where the pivot speed $\omega_{piv}$ is commanded based on the reaction wheel speed $\omega_{rw}$ and the commanded reaction wheel torque $\tau_{rw,comm}$ through gains $g_1$ and $g_2$, respectively. For coarse slews, a simple PI speed controller is used to command reaction wheel torque based on the yaw speed $\dot{\theta}_3$ from rate gyroscope measurements $\boldsymbol\omega_b$ projected to the outer frame:
\begin{eqnarray}
\tau_{rw,comm} = -k_P(\dot{\theta}_{3}-\dot{\theta}_{3,d})-k_I\int_0^t(\dot{\theta}_{3}-\dot{\theta}_{3,d})dt
\label{eq:rw_control}
\end{eqnarray}
Here, the desired yaw rate $\dot{\theta}_{3,d}$ follows a trapezoidal speed profile towards the target yaw angle $\theta_3$ with a constant acceleration of 0.5 deg/s\textsuperscript{2} on the rising edge, 0.1 deg/s\textsuperscript{2} on the falling edge, and a top speed of 4 deg/s. Furthermore, if controller coupling is ignored, it can be shown via the yaw dynamics that gains $k_P$ and $k_I$ can be chosen to asymptotically stabilize $\dot{\theta}_{3}$ while gains $g_1$ and $g_2$ can be chosen to asymptotically stabilize $\omega_{rw}$ \citep{RefWorks:85} according to
\begin{eqnarray}
I_{yaw}\ddot{\omega}_{rw}+g_2k_{ft}I_{yaw}\dot{\omega}_{rw}+g_1k_{ft}\omega_{rw} = g_1k_{ft}\omega_{rw,d}
\label{eq:rw_dyn_no_coupling}
\end{eqnarray}
where $I_{yaw}$ is the gondola inertia about the yaw axis and $k_{ft}$ is the torsional stiffness of the flight train. Note that these dynamics are a simplification of the full yaw dynamics, a discussion of which is beyond the scope of this work. Suffice it to say that if the fully coupled yaw dynamics are taken into account, it can be shown using a \textit{Routh stability analysis} \citep{RefWorks:93} that there are restrictions on the gains $g_1$ and $g_2$ based on the selection of gains $k_P$ and $k_I$ to maintain system stability (see Figure \ref{fig:yaw_stab_diag}). Consequently, for reasonable gain selection providing critically-to-near-over-damped speed control response from (\ref{eq:rw_control}), there is a clear upper limit in $g_2$ for a given $g_1$, which, in effect, limits the response time on stabilizing $\omega_{rw}$ to large time scales. This effect was observed empirically during BIT system development and for other balloon-borne payloads \citep{RefWorks:85}, where large pivot gains generally caused system instability. Despite this, the restriction on the controller gains in (\ref{eq:momentum_dump}) is deemed an acceptable limitation for coarse control.

The roll and pitch axes are controlled using two 5 N$\cdot$m Parker frameless DC motors per axis with 8-bit Advanced Motion Controls (AMC) PWM controllers for each (see Table \ref{table:actuator_char}). For the roll axis $\theta_1$, the gimbal position is servoed to the desired angle $\theta_{1,d}$ using feedback from absolute encoders and the PID control law
\begin{eqnarray}
\tau_{roll} &=& -k_{P,r}(\theta_1-\theta_{1,d})+\nonumber\\
&&...-k_{I,r}\int_0^t(\theta_1-\theta_{1,d})dt-k_{D,r}\dot{\theta}_{1}
\label{eq:roll_control}
\end{eqnarray}
where $\dot{\theta}_1$ is found from (\ref{eq:312_mapping_matrix}).

The pitch axis is identical to the roll axis except that it is doubly actuated by two coarse stepper motors through a 12:1 gear reducer to achieve a full pitch range of 20-57$^\circ$. For coarse moves, the stepper motors are commanded to a given position from a known ``home'' position ($\theta_2=19.2^\circ$) using a trapezoidal speed profile with constant acceleration of $\pm$0.5 deg/s\textsuperscript{2} on the rising/falling edges and a top speed of 1.0 deg/s to prevent damage to the telescope and optics. During the coarse motion, the pitch encoder is servoed to zero using the same type of PID controller as (\ref{eq:roll_control}). In order to reset the stepper motor count and encoder position, a coarse pitch synchronization routine is used, which zeros the pitch measurement at the home position.

\subsubsection{Fine pointing stabilization.}
\label{sss:fine_point_stab}

Once the gimbal position corresponding to the desired target location on the sky has been reached, the control algorithm switches to a fine pointing stabilization mode, where sky rotation and disturbances from pendulations are tracked using the rate gyroscopes and star cameras. To prevent unnecessary disturbances, coarse stepper motors on the pitch axis are in a locked state, which, as a result, limits the pitch gimbal angle to $\pm 10^\circ$ from the coarse pitch target. It should also be noted that the momentum dumping law given in (\ref{eq:momentum_dump}) continues to stabilize the reaction wheel speed during fine pointing stabilization.

From a controller perspective, in order to map torques from the body frame $\coord{b}$ to the gimbal axes, the following PID control law is used:
\begin{eqnarray}
\boldsymbol\tau_{app} &=& -(\mathbf{S}_\theta^T\mathbf{S}_\theta)^{-1}\left(\mathbf{K}_P\boldsymbol\theta_{err}+
\mathbf{K}_I\int_0^t\boldsymbol\theta_{err} dt\right)+\nonumber\\
&&...-\mathbf{S}_\theta^{-1}\mathbf{K}_D\boldsymbol\omega_b
\label{eq:fine_pid_control}
\end{eqnarray}
where the matrices $\mathbf{K}_P = \mathbf{K}_P^T > \mathbf{0}$, $\mathbf{K}_I = \mathbf{K}_I^T > \mathbf{0}$, and $\mathbf{K}_D = \mathbf{K}_D^T > \mathbf{0}$ are the proportional, integral, and derivative gains, respectively, and $\boldsymbol\omega_b$ is the angular velocity of the telescope in $\coord{b}$. Furthermore, the error term $\boldsymbol\theta_{err}$ is approximated by \citep{RefWorks:90}
\begin{eqnarray}
\boldsymbol\theta_{err}^\times = \mathbf{1}-\mathbf{C}_{bE}\mathbf{C}_{bE,d}^T
\label{eq:fine_error}
\end{eqnarray}
where $\mathbf{C}_{bE,d}$ is found from (\ref{eq:equat_to_horiz}) and (\ref{eq:horiz_to_gimbal}) given equatorial coordinates. This is a reasonable approximation since the fine pointing stabilization controller is only active once the coarse gimbal control has reached its target to within $<0.1^\circ$.

Given the mapping matrix $\mathbf{S}_\theta$ from (\ref{eq:312_mapping_matrix}), it can be shown \citep{RefWorks:33} that the torque in the body frame $\boldsymbol\tau_b$ is related to the applied gimbal torques $\boldsymbol\tau_{app}$ through the same mapping matrix 
\begin{eqnarray}
\boldsymbol\tau_b = \mathbf{S}_\theta\boldsymbol\tau_{app} = \mathbf{I}\dot{\boldsymbol\omega}_b + \boldsymbol\omega_b^\times\mathbf{I}\boldsymbol\omega_b
\label{eq:euler_dynamics}
\end{eqnarray}
where the right-hand side is Euler's equation for rotational dynamics \citep{RefWorks:90} ($\mathbf{I}$ is the inertia matrix for the telescope and inner frame). Using these facts and the Lyapunov function
\begin{eqnarray}
V = \frac{1}{2}\boldsymbol\omega_b^T\mathbf{I}\boldsymbol\omega_b + \frac{1}{2}\boldsymbol\theta_{err}^T\mathbf{K}_P\boldsymbol\theta_{err}
\label{eq:fine_lyapunov}
\end{eqnarray}
the PD components of the controller given in (\ref{eq:fine_pid_control}) ensure asymptotic stability as follows:
\begin{eqnarray}
\dot{V} &=&\boldsymbol\omega_b^T\mathbf{I}\dot{\boldsymbol\omega}_b+\boldsymbol\theta_{err}^T\mathbf{K}_P\dot{\boldsymbol\theta}
\label{eq:stab_proof}
\\
&=& \boldsymbol\omega_b^T\left(-\boldsymbol\omega_b^\times\mathbf{I}\boldsymbol\omega_b+ \mathbf{S}_\theta\boldsymbol\tau_{app}\right)+ \boldsymbol\theta_{err}^T\mathbf{K}_P\dot{\boldsymbol\theta}_{err}\nonumber\\
&=& \boldsymbol\omega_b^T\mathbf{S}_\theta\left(-(\mathbf{S}_\theta^T\mathbf{S}_\theta)^{-1}\mathbf{K}_P\boldsymbol\theta_{err}+ -\mathbf{S}_\theta^{-1}\mathbf{K}_D\boldsymbol\omega_b\right)+\nonumber\\
&&...+\boldsymbol\theta_{err}^T\mathbf{K}_P\dot{\boldsymbol\theta}_{err}\nonumber\\
&=& -\boldsymbol\omega_b^T\mathbf{S}_\theta^{-T}\mathbf{K}_P\boldsymbol\theta_{err}- \boldsymbol\omega_b^T\mathbf{K}_D\boldsymbol\omega_b + \boldsymbol\theta_{err}^T\mathbf{K}_P\dot{\boldsymbol\theta}_{err}\nonumber\\
&=& -\dot{\boldsymbol\theta}^T\mathbf{K}_P\boldsymbol\theta_{err} -\boldsymbol\omega_b^T\mathbf{K}_D\boldsymbol\omega_b +\boldsymbol\theta_{err}^T\mathbf{K}_P\dot{\boldsymbol\theta}_{err}\nonumber\\
&=&-\boldsymbol\omega_b^T\mathbf{K}_D\boldsymbol\omega_b \leq 0\nonumber
\end{eqnarray}
which holds for all $\boldsymbol\omega_b \in \mathbb{R}^3$ (it is assumed that reference $\mathbf{C}_{bE,d}$ is constant so that $\dot{\boldsymbol\theta} = \dot{\boldsymbol\theta}_{err}$) \citep{RefWorks:55}. Now, it is relatively straightforward to show that when the critical point $\boldsymbol\omega_b \dot{=} \mathbf{0}$ (which implies $\dot{\boldsymbol\omega}_b \dot{=} \mathbf{0}$) is substituted into the dynamics given by (\ref{eq:euler_dynamics}), the only configuration that satisfies the equation is $\boldsymbol\theta_{err} = \mathbf{0}$ if only the PD components of the controller in (\ref{eq:fine_pid_control}) are considerd. Thus, based on LaSalle's invariance principle \citep{Refworks:91}, this demonstrates asymptotic stability for $\boldsymbol\theta_{err} = \mathbf{0}$ under the approximation given in (\ref{eq:fine_error}). It should be noted that for the full PID controller, which includes the integral term, this stability is limited by the magnitude of $\mathbf{K}_I$ \citep{Refworks:94}. 
\section{Flight Performance}

The engineering results in the following subsections are taken from the 8 hour test flight of BIT from Timmins, Canada from September 18-19, 2015. Coarse and fine pointing stability and control for the telescope frame are given here and discussed in detail in the following section.

\subsection{Coarse attitude determination and control}
\label{ss:coarse_adcs_perf}

\begin{figure*}[h]
\centering
\includegraphics[width=1.01\textwidth]{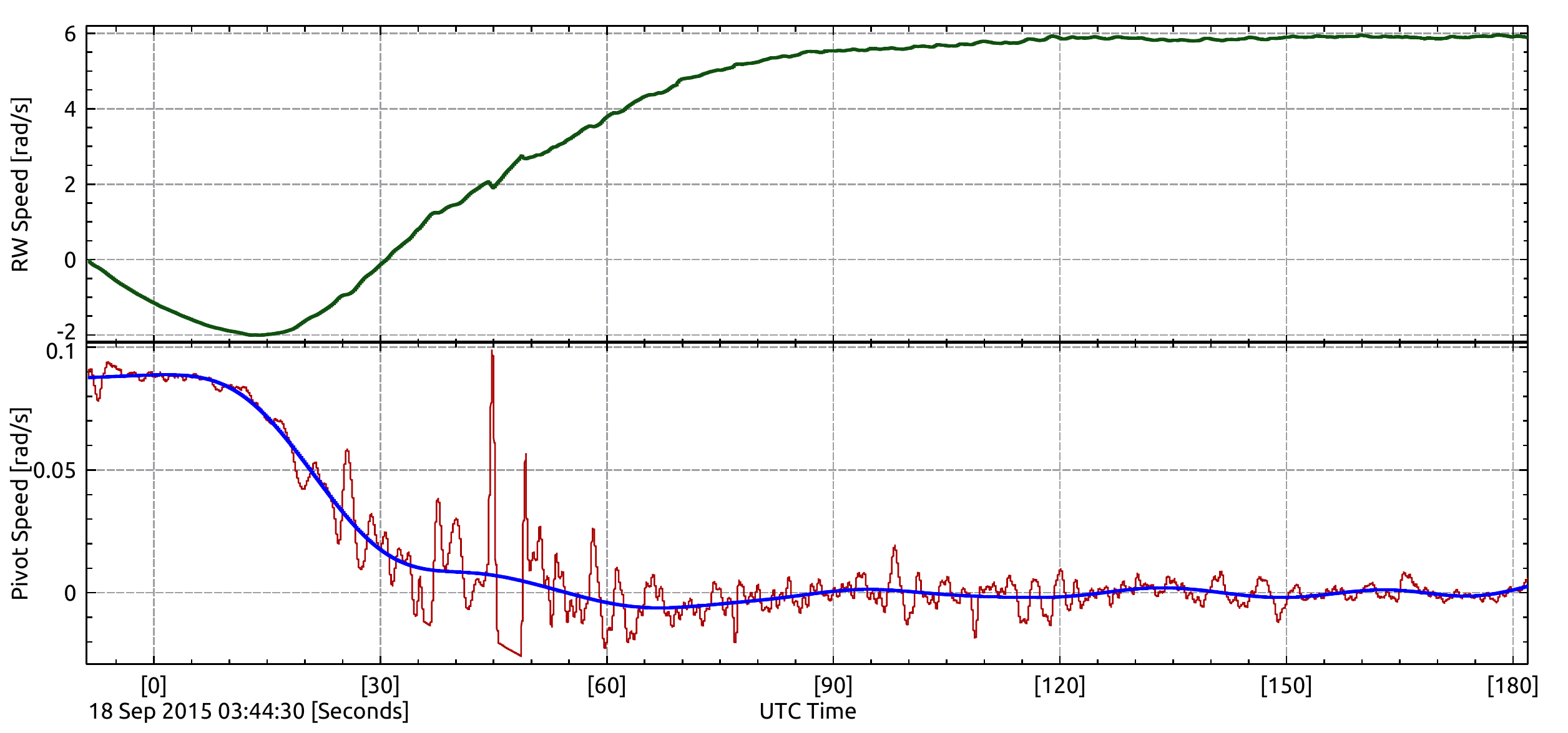}
\caption[numbered]{Reaction wheel (RW) response (top) over a period of three minutes while stabilizing on the sky with a set bias speed of 7 rad/s; commanded pivot speed (bottom, thick) tracks in the opposite direction to dump reaction wheel momentum, where noise on the pivot speed (bottom, thin) is due to contribution of the RW torque gain in (\ref{eq:momentum_dump}).}
\label{fig:pivot_rw_stab}
\end{figure*}

\begin{figure*}[h]
\centering
\includegraphics[width=1.01\textwidth]{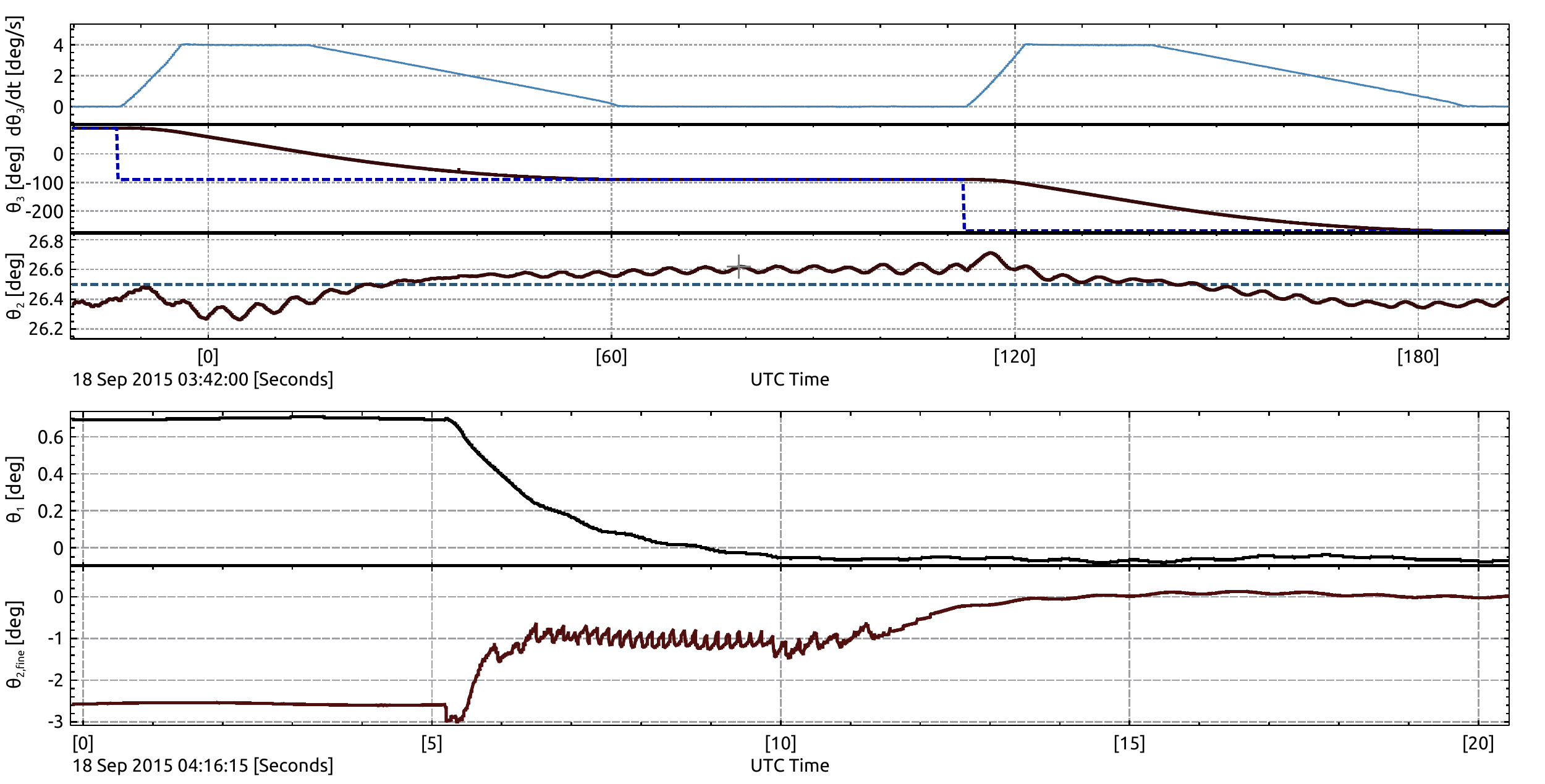}
\caption[numbered]{(Top) full $360^\circ$ yaw $\theta_3$ rotation through two $180^\circ$ slews during pre-flight; the yaw rate $d\theta_3/dt$ follows the prescribed trapezoidal profile to reach the commanded azimuth ($\theta_3$ dotted) within one minute with little overshoot in measured azimuth ($\theta_3$ solid); the pitch $\theta_2$ during slews varies by $0.1^\circ$ at $\sim 0.15$ Hz and $0.3^\circ$ at $\sim 0.01$ Hz from the highbay flight train; (bottom) coarse roll and pitch encoder measurements $\theta_1$ and $\theta_{2,fine}$ with a zero gimbal angle command over a 20 second period during pre-flight; coarse control acquires gimbal target to within < $0.1^\circ$ over 15 seconds, where low frequency periodic variations are due to pendulations.}
\label{fig:coarse_slews}
\end{figure*}

For coarse stability and pointing control involving large slews and coarse target acquisition, the primary flight results for gimbal control are given in Figures \ref{fig:pivot_rw_stab} and \ref{fig:coarse_slews}. As shown in Figure \ref{fig:pivot_rw_stab}, reaction wheel speed is stabilized over approximately three minutes to a steady state bias speed of about 6.1 rad/s using the yaw coarse control and momentum dumping scheme given in (\ref{eq:momentum_dump}) and (\ref{eq:rw_control}). With a commanded bias speed of 7 rad/s, the steady state error in reaction wheel speed of approximately 0.9 rad/s is due to the lack of an integral term in $\omega_{rw}$ for the pivot momentum dumping scheme given in (\ref{eq:momentum_dump}). Although an integral term could have corrected the steady state error, this was deemed unnecessary since reaction wheel bias momentum is present only to prevent the static friction region when controlling through the reaction wheel speed zero point; thus, the exact value of the bias momentum is irrelevant if it is sufficiently above zero. Furthermore, at +45 seconds, a change of $\sim 5^\circ$ in the desired yaw position was given, which accounts for the large spikes in commanded pivot speed. Despite this, however, it is clear that the reaction wheel speed stabilization using the momentum dumping scheme is unaffected.

When looking at the yaw performance of the gondola overall as shown in Figure \ref{fig:coarse_slews} (top), the response to the trapezoidal speed profile is clearly observed in the measured yaw rate $\dot{\theta}_3$, which provides a smooth transitional motion in azimuth even over large $180^\circ$ slews (note that here azimuth and elevation are approximately aligned with yaw and pitch, respectively). Since pendulations are not tracked during coarse slews, periodic variations in the elevation are observed, the magnitudes and frequencies of which are directly related to the highbay flight train. For the pitch and roll performance shown in Figure \ref{fig:coarse_slews} (bottom), the response to the requested gimbal angles for $\theta_1$ and $\theta_2$ shows critically damped behaviour to within < $0.1^\circ$, where periodic pedulations leak into the gimbal control through rate gyroscope measurements according to (\ref{eq:312_mapping_matrix}) and (\ref{eq:roll_control}) (i.e. there is no way to subtract the pitch and roll pendulations of the outer frame from the gimbal rates). During the coarse pitch change from between +6 and +11 seconds, the sawtooth pattern in the fine pitch encoder $\theta_{2,fine}$ measurement is directly associated with the coarse granularity of the pitch stepper motors cause by the limited resolution and precision in the stepper motor controllers. However, after the coarse pitch motion has terminated, the fine encoder measurements servo to an encoder angle of zero as desired.

\subsection{Fine attitude determination and control}
\label{ss:fine_adcs_perf}

\begin{figure*}[h]
\centering
\includegraphics[width=1.01\textwidth]{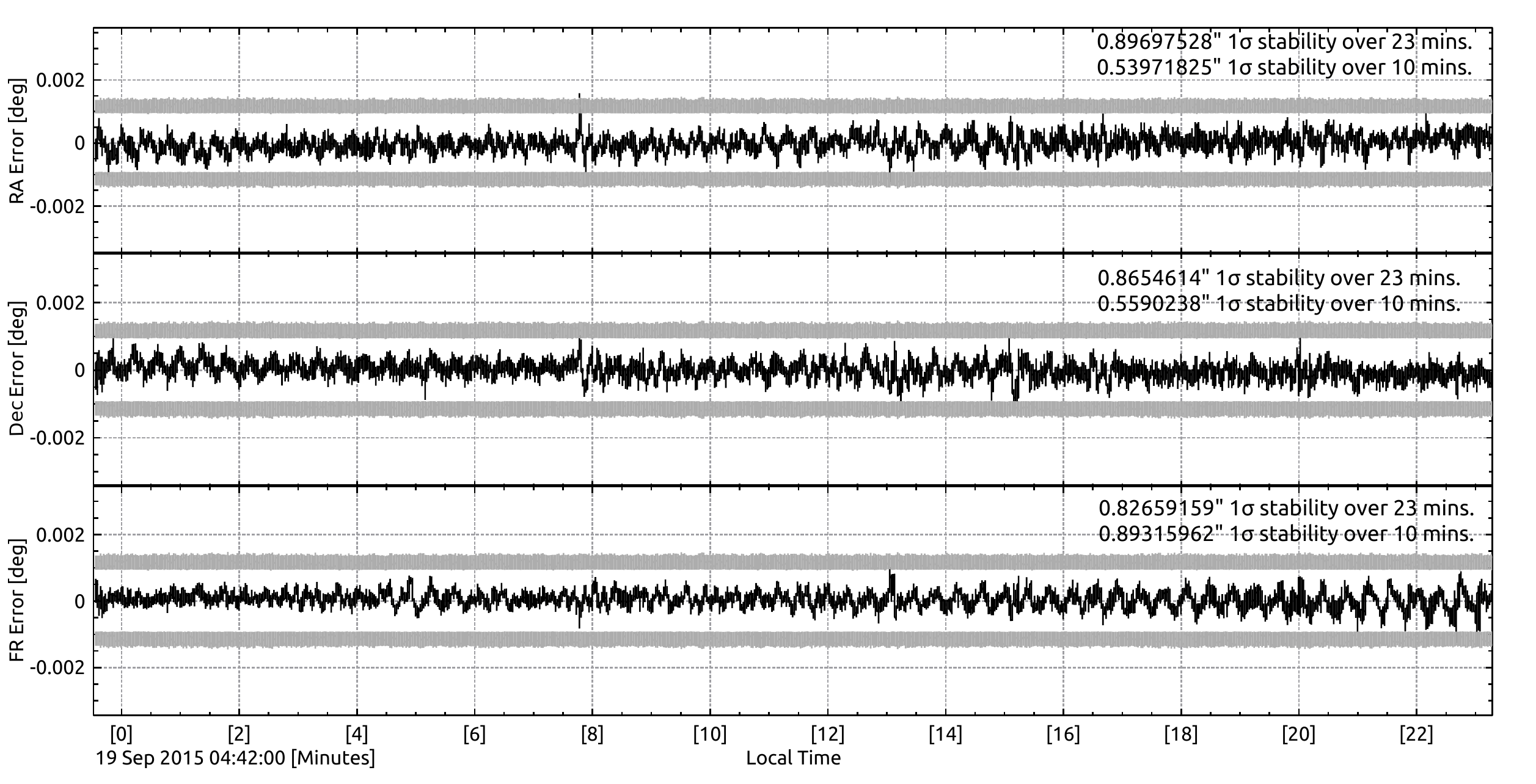}
\caption[numbered]{Pointing controller stability (dark) and $3\sigma$ envelope (light) for $RA$ (top), $Dec$ (middle), and $FR$ (bottom) over a 23 minute integration period; attitude estimation is consistent with the estimator covariance, where the variation in the $3\sigma$ envelope is due to asynchronous star camera measurements.}
\label{fig:pointing_stab_cons}
\end{figure*}

\begin{figure*}[h]
\centering
\includegraphics[width=1.01\textwidth]{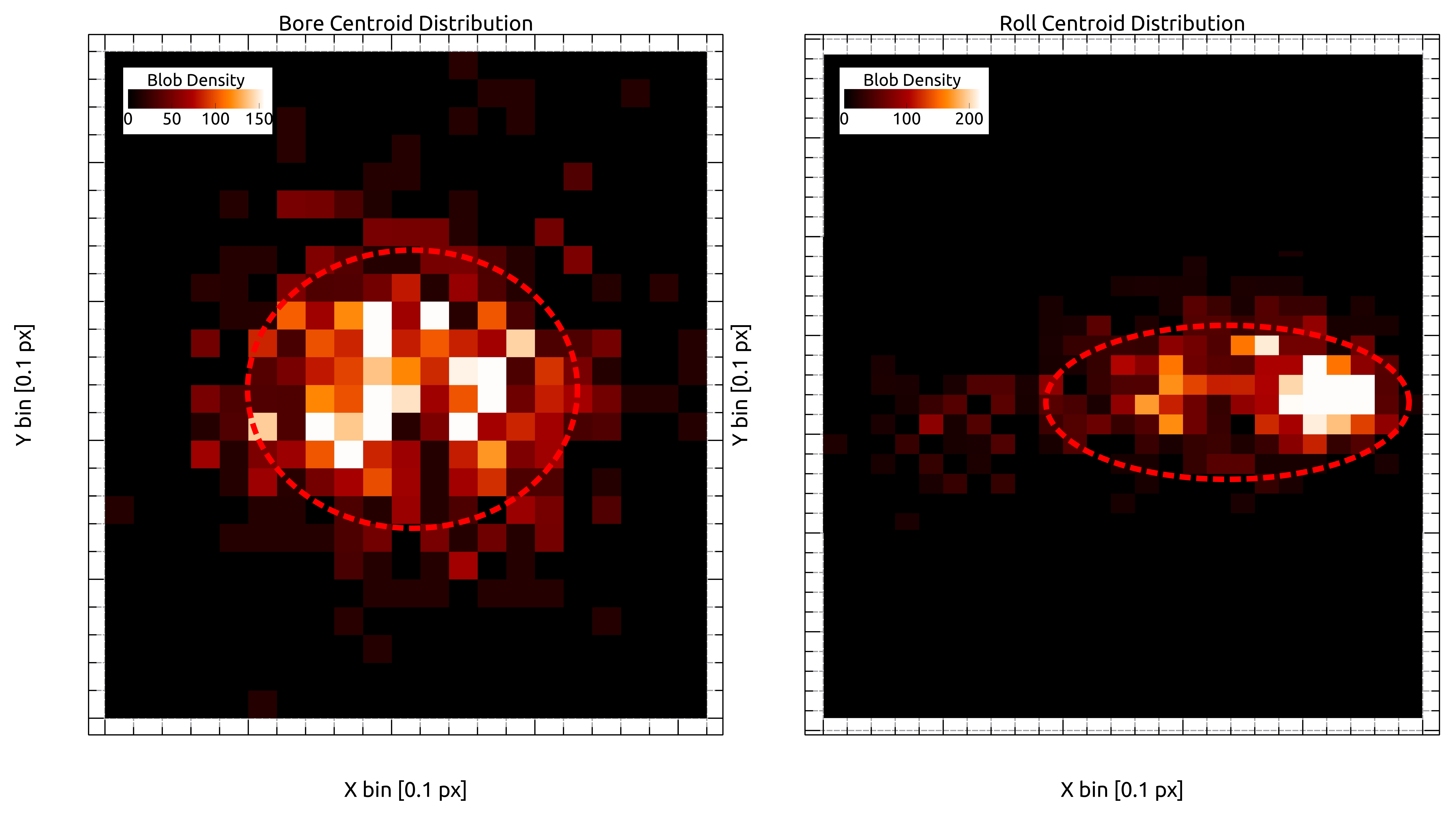}
\caption[numbered]{Bore star camera (left) and roll star camera (right) centroid locations over three minutes while tracking and stabilizing the telescope frame; $3\sigma$ ellipse (dashed) shows the spread of the centroids over the star camera focal planes, where the pixel scale for bore and roll are 2.3$^{\prime\prime}$/px and 4.3$^{\prime\prime}$/px, respectively; centroid locations artificially discretized to one-tenth of a pixel in software.}
\label{fig:roll_bore_centroid}
\end{figure*}

\begin{figure*}[h]
\centering
\includegraphics[width=1.01\textwidth]{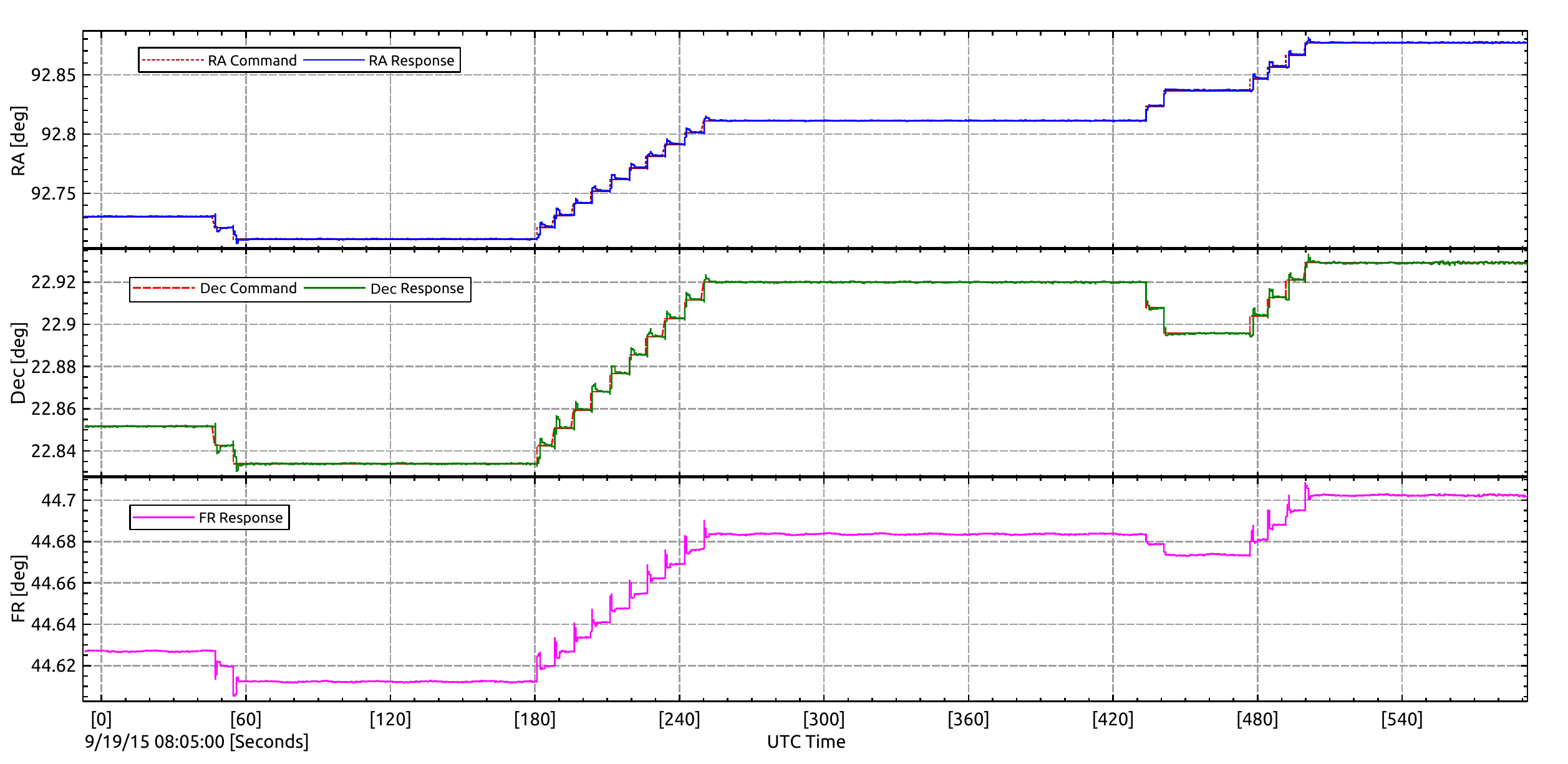}
\caption[numbered]{$RA$ (top), $Dec$ (middle), and $FR$ (bottom) responses to 1$^{\prime}$ step commands in $Az$ and $El$ while stabilizing on the sky over a 10 minute period; the high gain fine pointing control has some overshoot per step, but the settling time is on the order of < 2 seconds.}
\label{fig:pointing_delta_moves}
\end{figure*}

After stabilizing the telescope using coarse gimbal control, a series of fine pointing stabilization runs were performed in flight, during which the system was calibrated and tested for several star fields. Over the course of about 6.5 hours, two fine control runs were performed that stabilized the telescope at the sub-arcsecond level for more than an hour, where several shorter runs (on the order of 10-20 minutes) were performed for calibration and testing purposes. A segment of one of the hour long runs is given here as shown in Figures \ref{fig:pointing_stab_cons}-\ref{fig:pointing_delta_moves}. Note that all pointing results involving the telescope orientation $\mathbf{C}_{bE}$ are parametrized according to equatorial coordinates in frame $\coord{E}$ for readability.

\begin{figure*}[ht]
\centering
\includegraphics[width=0.9\textwidth]{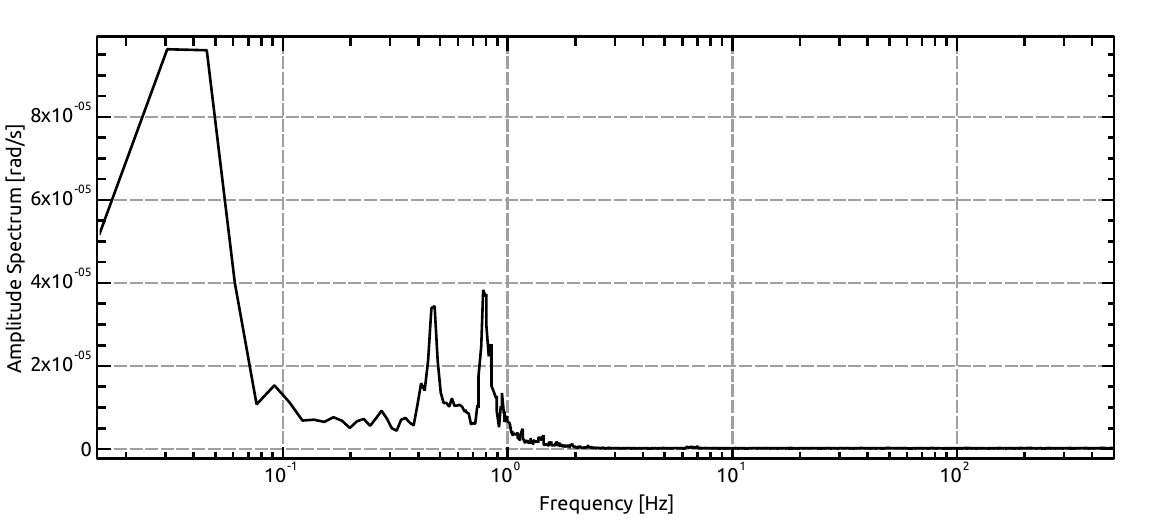}
\caption[numbered]{Amplitude spectrum for pitch rate gyroscope measurements in an uncontrolled state at a float altitude of 35 km; \changed{the dominant modes are clearly present in the low frequency regime (< 1 Hz);} peaks at 0.038 Hz and 0.78 Hz are due to compound pendulations the flight train about the balloon and the pivot; the central 0.47 Hz peak is likely due to the large communication electronics box used by the launch provider (CSA-CNES) located midway up the flight train.}
\label{fig:gyro_unc_amp_spect}
\end{figure*}

In Figure \ref{fig:pointing_stab_cons}, the results shown for attitude determination and control over the 23 minute period are twofold. From an attitude determination perspective, the restriction of the pointing error to the $3\sigma$ envelope for all three coordinates clearly demonstrates estimator consistency even over large time scales, which validates the confidence that the attitude determination scheme (i.e. EKF) has in the estimates that are produced \citep{RefWorks:40}. The thickness in the $3\sigma$ envelope is due to the increase in pointing solution variance from integrating rate gyroscope measurements between asynchronous star camera centroid measurements. Secondly from a control perspective, the pointing stability is clearly beyond the required specification of 1-2$^{\prime\prime}$, where total stability is 0.86$^{\prime\prime}$ (rms) over 23 minutes and 0.68$^{\prime\prime}$ (rms) over 10 minutes (averaged over all pointing measurement errors). Additionally, it was observed that pointing stability over longer time scales are consistently sub-arcsecond regardless of target on the sky, which demonstrates closed-loop, on-demand stabilization for arbitrary astronomical targets. Overall, compared to the magnitude of disturbances due to pendulations shown in Figure \ref{fig:gyro_unc_amp_spect}, an attenuation of approximately 100:1 was achieved with the BIT ADCS, where gains for controller given in (\ref{eq:fine_pid_control}) were only limited by rate gyroscope and star camera noise.

As shown in Figure \ref{fig:roll_bore_centroid}, the positions of the tracked star in the roll and bore star cameras are plotted over a three minute period of sub-arcsecond stabilization. It is clear that the bore centroids are well constrained in the $x$ and $y$ directions, where the $3\sigma$ ellipse is a good representation of the centroid spread. For the roll camera, however, it is evident that the spread is much wider in the $x$ direction than it is in the $y$ direction, which may indicate a slight drift in the roll centroids corresponding to a bias in the telescope roll component of the pointing solution. Likely causes for this drift and the possible effects it has on pointing stability are discussed in the following section. 

During the flight, a series of 1$^{\prime}$ moves in $Az$ and $El$ were performed to move bright stars from the telescope focal plane to the fine tracking camera used in the fine guidance system. The effects of these moves on the equatorial coordinates corresponding to the pointing solution are given in Figure \ref{fig:pointing_delta_moves}. From the results, it is clear that the high gain system causes some overshoot at each arcminute step, but overall system stability is assured based on (\ref{eq:stab_proof}). Furthermore, the sub-arcsecond stability of the pointing system between steps is recovered after a < 2 second settling time, which is an acceptable time scale for small changes in pointing targets between science camera integrations on the order of > 1 minute.

\section{Discussion}

For the coarse attitude determination and control overall, it is clear that the level of coarse stabilization during gimbal control and coarse pointing during large gimbal slews is adequate, where targets on the sky are acquired to within < $0.1^\circ$ of the measured attitude. Trapezoidal speed profiles perform to within acceptable limits, where defined constraints on angular acceleration and maximum speed are met. During the flight, however, it was observed that the component of the pointing solution in the azimuth direction would manifest large steady state errors of > $10^\circ$ with respect to $\coord{H}$ after long fine stabilization runs. One possible cause for this is that the magnetometer had not been calibrated fully due to time constraints during the flight such that its contribution to the attitude estimate was minimal (i.e. did not have a good inertial measurement of yaw to correct for drift while integrating using only rate gyroscopes). Although this would in principle have some effect on the absolute pointing error in this way, one would expect the drift in the pointing solution based on rate gyroscope measurements to be on the order of < $1^\circ$, given the high measurement frequency and relatively low $1/f$ noise characteristics (see Table \ref{table:sensor_char}).  

A more likely cause for this steady state yaw error is the residual magnitude of the pointing solution covariance after a fine stabilization run, during which the confidence in the pointing estimate is quite high. When mode switching occurred during the flight from fine to coarse (i.e. to acquire a new target after tracking), it was observed that the covariance of the attitude estimate increased, but not enough to maintain estimator consistency during coarse slews; thus, the state estimation scheme was overconfident in the attitude estimates when performing coarse slews immediately after fine stabilization. As a result, the attitude determination scheme would tend erroneously towards the previous fine stabilized attitude estimate, causing a large absolute error with respect to the current attitude estimate in $\coord{H}$ \citep{RefWorks:40}. Despite this problem in coarse pointing, steady state yaw errors were mitigated manually during flight by reacquiring target lock on the sky via lost-in-space from the star cameras. This can be corrected in future flights by either artificially increasing the covariance of the attitude estimate during mode switches from fine to coarse pointing or increasing the noise covariance associated with coarse gimbal measurement $\mathbf{C}_{bE,meas,coarse}$ (i.e. a covariance estimate reset).

From the fine attitude determination and control results, the target lock and fine stabilization performed exceptionally well, where the overall fine stability demonstrated was a factor of four times better than the required specification. Furthermore, the closed-loop step response of the system was well within acceptable limits, which was demonstrated over integration periods of more than an hour in duration. Although this is a positive result, it is evident from the centroiding data that there was a notable drift in the $x$ direction of the roll star camera focal plane. Taking into account the orientation of the roll camera with respect to the telescope frame, the pixel drift on the roll camera corresponds to a 0.25$^{\prime\prime}$/s drift  with respect to the sky along the axis parallel to the telescope boresight. Although the roll star camera measurements were contributing to the pointing solution, it was observed that the pixel measurements themselves were not being fully integrated; in other words, the roll camera centroids were clearly affecting the pointing solution through the reduction in the estimator covariance about the boresight axis, but the measurements used to correct the state were erroneous. It was discovered post-flight that there was a programming error that caused this behaviour, which can easily be corrected in subsequent flights. Despite the fact that this drift did introduce a slight coupling between $RA$ and $Dec$ over long integration periods, the overall stability of the telescope was not greatly affected. In fact, during the test flight, this was mitigated by manually adjusting the rate gyroscope bias about the telescope boresight axis to compensate for the roll drift.

Another aspect of the flight performance that is related to the pointing and stabilization of the telescope frame is the effect of the controller characteristics on image stabilization of the telescope focal plane down to 0.05$^{\prime\prime}$. Since the image stabilization control functions within the closed-loop regime of the telescope stabilization control, any residual disturbances or driven resonances from the telescope are propagated through to and perturb image stabilization. During the test flight, it was observed that there were significant higher frequency disturbances (> 5-10 Hz) seen on the focal plane of the telescope, which the image stabilization controller could not fully compensate for. Although there were other factors contributing to this lack of bandwidth in image stabilization (the discussion of which is beyond the scope of this paper), it is likely that this effect could be mitigated to a certain degree by reducing the gains on the telescope stabilization controller. In this way, the pointing stability of the telescope would be somewhat diminished, but with the possible benefit of reducing the frequency of the disturbances to within the bandwidth of the image stabilization control. For future flights, this trade-off can be further explored to see exactly how much control bandwidth can be gained from an image stabilization perspective.

\section{Conclusions}

Overall, the attitude determination and control systems for BIT performed adequately during the 2015 test flight, during which both the pointing and stabilization specifications were demonstrated thoroughly. Some improvements in handling mode switching from fine stabilization to coarse target acquisition can be made such that manual corrections for absolute steady state errors are no longer necessary. Furthermore, improving fine stability about the telescope bore axis can certainly improve the overall performance of the fine stabilization control such that long time-scale drifts are corrected for by design. In addition to this, a reduction in telescope stabilization gains may improve image stabilization, but this effect is likely marginal since most of the major improvements on the image stabilization front are unrelated to telescope control. 

For the proposed ULDB flight from New Zealand in 2018, a number of modifications must be made to the attitude determination and control systems in order to improve robustness, reliability, and long-term use. First of all, the overall autonomy of the system must be increased such that minimal human intervention is required to acquire and lock on to astronomical targets, where target prioritization and scheduling is done independently on flight hardware. Similarly, the ability of the pointing systems to self-calibrate attitude sensors and actuators is an absolute requirement for future missions due to limited communication windows with the payload for flights as long as three months. Lastly, measures must be taken to improve the robustness of the attitude determination and control systems against erroneous states, which is critical to demonstrating the capabilities of BIT as a facility class instrument and for ensuring reliability and consistency of the system for the upcoming flights.

%%%%Named reference model and we can use the same as numbered reference
%%%%giving option by using natbib package

\bibliography{allrefs}

\clearpage
\newpage

\begin{table}
\caption{BIT attitude sensor characteristics}
\label{table:sensor_char}
\begin{tabular}{*{1}{l}*{3}{c}r}
	\toprule
	Sensor Description & Readout Frequency (Hz) & Resolution & Noise Figure	\\
	\midrule
	Fibre optic rate gyroscope & 1000\textsuperscript{a} & $4.768\cdot 10^{-4}$ deg/s & $2.2\cdot 10^{-4}$ deg/(s$\cdot\sqrt{\mbox{Hz}}$)\\
	Absolute optical encoder & 100 & $5.49\cdot 10^{-3}$ deg & - \\
	3-axis magnetometer & 20 & $6.7\cdot 10^{-5}$ Gs & $2.0\cdot 10^{-4}$ Gs\\
	Coarse elevation stepper & 10 & $9.374\cdot 10^{-3}$ deg & -\\
	Bore star camera & 3 & $0.23^{\prime\prime}$ centroids & $5.75\cdot {10^{-4}}^{\prime\prime}$/s\textsuperscript{b}\\
	Roll star camera & 3 & $0.46^{\prime\prime}$ centroids & $5.75\cdot {10^{-4}}^{\prime\prime}$/s\textsuperscript{b}\\
	\bottomrule
\end{tabular}
\\
\small{\textsuperscript{a} Asynchronous serial ($\pm 5\%$) remapped to synchronous 1000 Hz via Akima interpolation \cite{RefWorks:92}}
\\
\small{\textsuperscript{b} Sky equivalent read noise}
\end{table}
\begin{table}
\caption{BIT actuator characteristics}
\label{table:actuator_char}
\begin{tabular}{*{1}{l}*{2}{c}r}
	\toprule
	Actuator Description & Control Input & Characteristics\\
	\midrule
	Reaction wheel - frameless DC motor & 16-bit analog & 15 N$\cdot$m max. torque; 3600 lines/rev encoder feedback\\
	Pitch/roll - frameless DC motor $\times$ 4& 8-bit PWM\textsuperscript{a} & 5.0 N$\cdot$m max. torque; 3-phase Hall sensor feedback\\
	Pivot - 2-phase stepper motor & pulse step/direction & 0.018 deg/step\textsuperscript{b}; 2-256 \si\micro step/step; 0.44 N$\cdot$m hold \\
	Pitch - 2-phase stepper motor $\times$ 2 & pulse step/direction & 0.15 deg/step\textsuperscript{c}; 16 \si\micro step/step; 0.51 N$\cdot$m hold\\
	\bottomrule
\end{tabular}
\\
\small{\textsuperscript{a} Pulse-Width Modulation}\\
\small{\textsuperscript{b} 1.8 deg/step motor through a 100:1 gear reducer}\\
\small{\textsuperscript{c} 1.8 deg/step motor through a 12:1 gear reducer}
\end{table}

%%%%%%%%DOCUMENT END HERE
\end{document}